\documentstyle[12pt]{article}
\renewcommand
\baselinestretch{1.3}

\begin{document}


\hfill July 1993

\vspace*{3mm}

\begin{center}

{\LARGE \bf
Effective potential for the conformal sector of quantum
gravity with torsion}

\vspace{4mm}

\renewcommand
\baselinestretch{0.8}
\medskip

{\sc E. Elizalde}\footnote{On leave of absence from and
permanent address: Department E.C.M., Faculty of Physics,
Barcelona University, Diagonal 647, 08028 Barcelona, Spain;
e-mail: eli @ ebubecm1.bitnet}
\\ {\it Division of Applied Mechanics, The Norwegian Institute
of Technology, \\ and Department of Theoretical Physics,
University of Trondheim, \\ N-7034 Trondheim, Norway} \\  and
\\
{\sc S.D. Odintsov} \\ {\it
Tomsk Pedagogical Institute, 634041 Tomsk, Russia}

\renewcommand
\baselinestretch{1.4}

\vspace{5mm}

{\bf Abstract}

\end{center}

The effective potential which describes the conformal dynamics
of quantum gravity with torsion is discussed. The phase
transitions induced by the combination of torsion and
curvature are investigated. The mechanism for fixing the
vacuum expectation values of the metric and torsion is
presented.

\vspace{4mm}

\newpage

\noindent 1. \ The concept of effective potential plays a very
important role
in modern particle physics [1].  The number and the variety of
applications of effective pootentials is growing continuously
[1] and even the quite complicated two-loop effective
potential for the standard model could be found [2], summing
also the leading logarithms.

{}From another point of view, the scalar field effective
potential has
many application to early universe considerations. For
instance, such a potential is required for the analysis of
the inflationary universe (see [3], for a general review),
where curvature effects are of the essence. It is also rather
well-known [4], that the effective potential in curved
spacetime shows very interesting features, such as dynamical
symmetry breaking induced by curvature [4] and gravitational
phase transitions (for a general review see [5]).

Recently it has been found that the effective potential
concept may be important also for fixing up the vacuum
expectation value in quantum gravity. One of such scenarios is
based on the conformal dynamics of quantum gravity, as
developed in papers [6] very recently. The conformal
dynamics of quantum gravity are induced by the conformal
anomaly and can be reduced to the description of infrared
quantum gravity, i.e., of quantum gravity at very large scales
[6].

The conformal dynamics of quantum gravity with torsion have
been further discussed in refs. [7], which develop the
approach of ref. [6]. The inclusion of torsion in the
discussion seems quite reasonable nowadays, since we certainly
know that torsion appears in string theory in a natural way
(as an axion). Furthermore, there is an interest for torsion
in some other contexts, steming from the search of the
so-called fifth force.

In a recent work, [8], an attempt has been made at studying
the phase structure of the effective potential corresponding
to the conformal factor, in the effective theory of infrared
quantum gravity [6]. For instance, the curvature-induced (or
gravitational) phase transitions which take place  in such a
situation can be considered (in some context) as the analog of
the $c=1$ phase transition of non-critical string theory.

In the present work we study the phase structure of the
conformal-factor effective potential for quantum gravity with
torsion. The general expression describing the conformal
dynamics of quantum gravity with torsion in a curved fiducial
background is obtained. The one-loop effective potential for
the conformal factor (up to terms linear on the curvature and
up to second order in torsion) is also calculated. Then, the
phase transitions induced by the combination of background
torsion and curvature are investigated. As a consequence, the
relevant mechanism for fixing the vacuum expectation values of
the metric and torsion, at the critical point corresponding to
the phase transition, is given. Notice that a similar idea
consisting in fixing the metric at the minimum of the efective
potential corresponding to the conformal factor ---for the
case of a flat background in the approach [9]--- have been
discussed in ref. [10] recently, within the framework of
multiplicatively renormalizable $R^2$-gravity [12].
\bigskip

\noindent 2. \ Let us start from the trace anomaly for a free
conformal
invariant theory in curved spacetime with torsion (see refs.
[5,11] for details)
\begin{eqnarray}
T_\mu^{\ \mu} &=& b C^2_{\mu\nu \alpha\beta} + b' \left( G-
\frac{2}{3} \Box R \right) + \left[ b'' +\frac{2}{3}
(b+b')\right] \Box R + a_1 F_{\mu\nu}^2 \nonumber \\
&&+ a_2 ( S_\mu S^\mu )^2 + a_3 \Box ( S_\mu S^\mu )  + a_4
\nabla_\mu ( S_\nu \nabla^\nu S^\mu - S^\mu \nabla_\nu S^\nu
),
\end{eqnarray}
where $G$ is the Gauss-Bonnet invariant, $F_{\mu\nu} =
\nabla_\mu S_\nu - \nabla_\nu S_\mu$, $b$, $b'$ and $b''$ are
known (see, for example, [11]), and where the coefficients
$a_i$ relevant for non-zero torsion are given by [5]
\begin{eqnarray}
&& a_1 =- \frac{2}{3(4\pi)^2} \Sigma \eta^2, \ \ \ \ a_2=
\frac{1}{2(4\pi)^2} \Sigma \zeta^2, \nonumber \\
&& a_3 = \frac{1}{3(4\pi)^2} \Sigma \left( 2\eta^2-
\frac{1}{2}
\zeta^2 \right), \ \ \ \ a_4= - \frac{2}{3(4\pi)^2} \Sigma
\eta^2.
\end{eqnarray}

Notice that the coupling constants that appear in (2) come
from the free conformally invariant theory (spinors and
scalars) in torsionful spacetime:
\begin{eqnarray}
S_0 &=& \frac{1}{2} \int d^4x \, \sqrt{-g} \left( g^{\alpha
\beta}  \partial_\alpha \varphi \partial_\beta \varphi +
\frac{1}{6} R\varphi^2 + \zeta_1  S_\mu S^\mu \varphi^2
\right),
\nonumber \\
S_{1/2} &=& i \int d^4x \, \sqrt{-g} \left[ \bar{\psi} \left(
\gamma^\mu  \nabla_\mu -\eta   \gamma_5 \gamma^\mu S-\mu
\right)\psi \right].
\end{eqnarray}
These theories (3) are conformally invariant, for any
$\zeta_1$
and $\eta$, and the minimal coupling corresponds to   $\zeta_1
=0$ and $\eta =1/8$. Notice that vectors do not interact at
all minimally with torsion (see, for example, ref. [5]).

A further remark has to do with quantum gravity effects
themselves. Generally speaking, the coefficients of the trace
anomaly (1) contain also a contribution from spin 2 modes.
Such contribution to $b$ and $b''$ (coming from the Einstein
or Weyl gravity action) have been calculated explicitly in the
last one of refs. [6]. It is to be expected that some new
contributions from gravity to $a_1$, $\ldots a_4$ will appear,
because we now consider quantum gravity with torsion. However,
it turns out that quantum gravity contributions to   $a_1$,
$\ldots a_4$ remain absent, owing to the fact that torsion
shows up without derivatives (at least in the Einstein-Cartan
theory). Hence, the torsion field looks non-dynamical. Of
course, a direct calculation is necessary in order to prove
this statement. Moreover, the situation will certainly change
in higher derivative quantum gravity with torsion, in which
case torsion is actually dynamical, already at the classical
level.

As the next step of our procedure, we choose the conformal
parametrization
\begin{equation}
g_{\mu\nu} = e^{2\sigma (x)} \bar{g}_{\mu\nu}, \ \ \ \ S_\mu
=\bar{S}_\mu,
\end{equation}
where $\sigma$ is the conformal factor, $ \bar{g}_{\mu\nu}$ is
a fixed fiducial metric, and $\bar{S}_\mu$  is an arbitrary
torsion background. Then one can integrate over the trace
anomaly in order to get the trace-anomaly-induced action
$S_{anom}$ (see [5-7] for details).

Adding the classical gravity action
\begin{equation}
S_{cl} = \frac{1}{2k}  \int d^4x \, \sqrt{-g} \left( R +
hS_\mu S^\mu -2\Lambda  \right)
\end{equation}
in the parametrization (4) to $S_{anom}$, we get the total
effective action which describes the quantum conformal factor
dynamics
\begin{eqnarray}
S_{eff} &=& S_{anom} +  S_{cl} =  \int d^4x \, \sqrt{-g}
\left\{ - \frac{\theta^2}{(4\pi )^2} \sigma \Box^2 \sigma +
\sigma
\left[ 2 \left( \zeta -  \frac{\theta^2}{(4\pi )^2}  \right)
R^{\mu\nu} \nabla_\mu  \nabla_\nu \right. \right. \nonumber \\
&& \left. - \left( \zeta -
\frac{2\theta^2}{3(4\pi )^2}  \right) R \Box
-  \frac{1}{3(4\pi )^2} \, \frac{\theta^2}{(4\pi )^2}
(\nabla_\mu R)  \nabla^\mu \right] \sigma \nonumber \\
&&- \zeta \left[2
\alpha
(\nabla_\mu \sigma) (\nabla^\mu \sigma) \Box \sigma + \alpha^2
\left( (\nabla_\mu \sigma) (\nabla^\mu \sigma) \right)^2
\right] \\
&& + \frac{\gamma}{6 \alpha^2} e^{2\alpha \sigma} R + \left[
b'' +\frac{2}{3} (b+b')\right]  R  (\nabla_\mu \sigma)
(\nabla^\mu \sigma)  + \gamma \, e^{2\alpha \sigma}
(\nabla_\mu \sigma) (\nabla^\mu \sigma) - \frac{\lambda}{
\alpha^2} e^{4\alpha \sigma} \nonumber \\
&& + \left. \frac{h}{2k\alpha^2} e^{2\alpha \sigma} S^2 +
\left( a_3 +\frac{a_4}{2}\right) S^2 (\nabla_\mu \sigma)
(\nabla^\mu \sigma)  + a_4 S^\mu S^\nu  (\nabla_\mu \sigma)
(\nabla_\nu \sigma) \right\}, \nonumber
\end{eqnarray}
where $\theta^2/(4\pi )^2 =2b +3b''$, $\zeta =2b +2b' +3b''$,
$\gamma = 3/k$ and $\lambda = \Lambda /k$, and where the
transformations $\sigma \rightarrow \alpha \sigma $ and
$S_{eff} \rightarrow \alpha^{-2} S_{eff}$ have been performed.
One should notice that, when calculating $S_{anom}$ we have
dropped the
$\sigma$-independent terms and also the terms linear on
$\sigma$  (as is usually done in quantum field theory).
Moreover, the bar over $g^{\mu\nu}$ and $S_\mu$ has been
omitted in (6) ---as will be done in what follows. The only
quantum field in Eq. (6) is $\sigma$.

The following remark is in order. In two dimensions, a
conformal parametrization, such as (4), completely fixes the
metric. Hence, the metric can be chosen to be flat, for
simplicity. On the contrary, in four dimensions this is of
course not the case, and we need to consider the curved
fiducial background $\bar{g}_{\mu\nu}$. As we will see, this
makes things rather non-trivial and leads to a very
complicated effective potential.
\bigskip

\noindent 3. \ The one-loop $\beta$-functions in the purely
gravitational sector of the theory defined by the action (6)
have been calculated in refs. [6], and in the case of a
torsionful sector, in refs. [7]. In what follows, we will
investigate the theory given by (6) around the infrared stable
fixed point $\zeta =0$, which presumably describes $4d$
quantum gravity at large distances (i.e., infrared quantum
gravity [6]). As is clear, the action (6) becomes much simpler
at  $\zeta =0$.
Let us now study the effective potential of the composite
field $\Phi = e^{\alpha \sigma}$ (one should have in mind that
the classical scaling dimension of $\Phi$ is $1-\alpha$).

As first example, we shall choose $\bar{S}_\mu =0$ and
$\bar{g}_{\mu\nu}= \eta_{\mu\nu}$. Then, one finds that the
Coleman-Weinberg like effective potential is given by
\begin{equation}
V^{(1)} (\Phi) = \frac{\lambda}{\alpha^2} \Phi^4 +\frac{1}{2}
\left[ \frac{\gamma^2 (4 \pi )^2}{2 \theta^4} -
\frac{8\lambda}{\theta^2} \right] \Phi^4 \left( \ln
\frac{\Phi^2}{\mu^2} -\frac{25}{6} \right).
\end{equation}
By simple inspection of the form of this potential we already
see that its classical minimum corresponds to the singularity
$\Phi=0$, in terms of the original metric.

As a result of the Coleman-Weinberg symmetry breaking ($\mu^2
= e^{2 \alpha \sigma_0}$),
\begin{equation}
\frac{\Phi}{\mu^2} =   \exp \left[ \frac{11}{3} -
\frac{2\lambda}{\alpha^2 \left(\frac{\gamma^2 (4 \pi )^2}{2
\theta^4} - \frac{8\lambda}{\theta^2} \right)} \right].
\end{equation}
We thus find that the singularity has been avoided, because
now a non-singular value of $\Phi$ corresponds to the ground
state (minimum) of the potential. Notice that similar type of
(logarithmic) corrections to the conformal-field effective
potential inquantum $R^2$-gravity have been discussed in ref.
[10]. The appearence of logarithmic corrections which,
generally speaking, destroy the general covariance in terms of
the original metric $g_{\mu\nu}$ has been explained there as
being a result of the regularization procedure. Some
indications were found also pointing out at the fact that the
physical metric to be used should be actually
$\bar{g}_{\mu\nu}$, and not $g_{\mu\nu}$.

As the next step, we will now calculate the effective
potential for non-zero background curvature and torsion. For
such a calculation we shall adopt the approach  used in [14]
(see also [5]), in which we only consider linear dependences
on invariants of the curvature and torsion, namely we suppose
that $\Phi^2 >> |R|$ and $\Phi^2 >> S^2$ (here $R$ and $S$ are
the fiducial curvature and torsion, respectively).  Using the
techniques of ref. [14] and the explicit form of the one-loop
$\beta$-functions [6,7] of the theory (6), we obtain the
following efective potential
\begin{eqnarray}
V^{(1)} (\Phi)& =& \frac{\lambda}{\alpha^2} \Phi^4
+\frac{1}{2} \left[ \frac{\gamma^2 (4 \pi )^2}{2 \theta^4} -
\frac{8\lambda}{\theta^2} \right] \Phi^4 \left( \ln
\frac{\Phi^2}{\mu^2} -\frac{25}{6} \right) \nonumber \\
&&- \frac{h \gamma}{6 \alpha^2} \Phi^2 S^2 +  \left[
\frac{\gamma (4 \pi )^2}{2 \theta^4}\left( a_3 +\frac{3}{4}
a_4 \right) -\frac{h\gamma}{6\theta^2} \right] \Phi^2 S^2
\left( \ln \frac{\Phi^2}{\mu^2} -3 \right) \nonumber \\
&&- \frac{ \gamma}{6 \alpha^2} \Phi^2 R +
\frac{\gamma}{6\theta^2} \, \Phi^2 R \left( \ln
\frac{\Phi^2}{\mu^2} -3 \right),
\label{v1f}
\end{eqnarray}
where Coleman-Weinberg normalization conditions have been used
(in the absence of torsion, gravitational phase transitions
have been discussed in ref. [8]).
\bigskip

\noindent 4. \ The rest of the work is devoted to the
investigation
of phase transitions induced by curvature and torsion. It is
interesting to notice that the effective potential (9) should
be relevant also in the discussion of the problem of the
cosmological constant [13].

Starting from Eq. (\ref{v1f}) and viewing the potential
$V^{(1)}$ as a function of $\Phi^2$, $R$ and $S^2$, it is not
difficult to see that a critical point satisfying the
conditions of our approximation (in particular, of course,
different from the trivial one at 0) can be obtained only in
the case that the following equation is satisfied:
\begin{equation}
 a_3 +\frac{3}{4} a_4 =\frac{h\theta^2}{24\pi^2},
\label{co1}
\end{equation}
and also that $\theta^2 > \alpha^2 \ln (7/6)$.
Provided that this is the case, than a whole plane of critical
points is obtained at
\begin{eqnarray}
&&\frac{\Phi_{cr}^2}{\mu^2} =3+ e^{\theta^2/\alpha^2}, \\
&& \frac{h\, R_{cr} + S_{cr}^2}{\mu^2} = -
e^{\theta^2/\alpha^2} \left\{
\frac{2\lambda}{\alpha^2} + 4 \left(
\frac{2\pi^2\gamma^2}{\theta^4} - \frac{\gamma}{\theta^2}
\right)  \left[ 2 \ln \left(  e^{\theta^2/\alpha^2} -7/6
\right) + \frac{  e^{\theta^2/\alpha^2} +3}{
e^{\theta^2/\alpha^2} -7/6} \right] \right\}, \nonumber
\end{eqnarray}
all of them with the same value for the potential:
\begin{equation}
V^{(1)}_{cr} =\mu^4 \left( 3+ e^{\theta^2/\alpha^2} \right)^2
\left[ \frac{\lambda}{\alpha^2} + 4 \left(
\frac{2\pi^2\gamma^2}{\theta^4} - \frac{\gamma}{\theta^2}
\right)  \ln \left(  e^{\theta^2/\alpha^2} -7/6 \right)
\right].
\end{equation}
The conditions of our approximation are satisfied provided
that
\begin{equation}
\theta^2 >> \frac{2\pi^2 \gamma^2}{\lambda}.
\end{equation}

As second case, we will now investigate the situation in which
the
$\beta$-functions vanish [6] ---hence, the trace of the
energy-momentum tensor for the $\sigma$-field sector vanishes
as well, as in $2d$ gravity [15]. The conditions for the
vanishing of the exact beta functions for $\gamma$ and
$\lambda$ are:
\begin{equation}
\alpha_\pm = \frac{\theta^2}{2} \left( 1 \pm \sqrt{1 -
4/\theta^2} \right), \ \ \ \ \ \frac{\lambda}{\gamma^2} =
\frac{2\pi^2}{\theta^2} \left( 1 + \frac{4\alpha^2}{\theta^2}
+ \frac{6\alpha^4}{\theta^4} \right).
\label{vb}
\end{equation}
The first equation (\ref{vb}) determines, at the fixed point,
the anomalous scaling dimension $\alpha$ in terms of the
central charge $\theta^2$. Using the conditions  (\ref{vb}),
we see that the one-loop $\beta$-function for $h$ vanishes at
[7]
\begin{equation}
a_3 = -\frac{3}{4} \, a_4.
\label{vb2}
\end{equation}
This corresponds to putting $\eta^2 = \zeta_1^2$.
Using these conditions, (\ref{vb}) and   (\ref{vb2}), in the
effective potential (9), we can investigate again the
possiblity of a phase transition. It is easy to see,
from the analysis carried out before for the general case,
that here no critical point is obtained, due to the fact that
the necessary condition (\ref{co1}) is not satisfied
(unless $h$ or $\theta$ is equal to zero, what would be
unphysical). However, in the extreme case of very small $h$ we
again obtain a critical point, which must necessarily be
considered as driven by the torsion $S^2$, and which is
obtained as a particular case of the first analysis above (the
same equations are valid, just setting $h=0$).
\bigskip

\noindent 5. \ In conclusion, we have discussed in this paper
the
phase structure of the effective potential for the conformal
sector of quantum gravity with torsion. A concrete mechanism
that defines the vacuum expectation values of the metric and
torsion as the ones which yield a corresponding minimum of the
effective potential has been presented. For instance, in the
simplest case when $\bar{g}_{\mu\nu} =\eta_{\mu\nu}$ and
$\bar{S}_\mu =0$ the expectation value of the conformal factor
(8) defines the expectation value of the original metric
$g_{\mu\nu} =D^{2/\alpha} \eta_{\mu\nu}$.

It is known that for $\Lambda =0$ $1/\kappa =3...?$ there
exist classical time dependent solutions for the conformal
factor, in theories of the De Sitter type. These theories may
be understood, for example, in terms of four dimensional
conformal gravity, as corresponding to inflationary universes.
Hence, it would be interesting to study the effective action
for $\sigma$. In fact, the time dependent solutions of this
effective action may find important applications in
inflationary cosmology.

\vspace{5mm}

\noindent{\large \bf Acknowledgments}

E.~Elizalde is grateful to Prof.~I.~Brevik and
Prof.~K.~Olaussen,
and to
Prof.~L.~Brink for the hospitality extended to him at the
Universities
of Trondheim and G\"{o}teborg, respectively.
This work has been  supported by DGICYT (Spain) and by CIRIT
(Generalitat de Catalunya).

\newpage


\newpage


\begin{thebibliography}{99}

\bibitem{} S. Coleman  and E. Weinberg,  Phys. Rev. {\bf D7}
(1973) 1888; M.B. Einhorn and D.R.T. Jones, Nucl. Phys. {\bf
B211} (1983) 29;  {\bf B230} (1983) 261; G.B. West,
Phys.Rev. {\bf D27} (1983) 1402; K. Yamagishi, Nucl. Phys.
{\bf B216} (1983) 503; M. Sher, Phys. Rep. {\bf 179} (1989)
274.

\bibitem{} C. Ford, I. Jack and D.R.T. Jones, Nucl. Phys. {bf
B387} (1992) 373;  C. Ford, D.R.T. Jones, P.W. Stephenson and
M.B. Einhorn, Nucl. Phys. {bf B395} (1993) 17.

\bibitem{} E.W. Kolb and M.S. Turner, {\it The Early
Universe},
Addison-Wesley, 1990; A.D. Linde, {\it Particle Physics and
Inflationary
Cosmology}, Contemporary Concepts in Physics, Harwood
Academic, New
York, 1990

\bibitem{} G.M. Shore, Ann. Phys. NY {\bf 128} (1980) 376.

\bibitem{} I.L. Buchbinder, S.D. Odintsov and I.L. Shapiro,
{\sl
Effective Action in Quantum Gravity}, IOP Publishing, Bristol
and
Philadelphia, 1992.

\bibitem{} I. Antoniadis and E. Mottola, Phys. Rev. {\bf D45}
(1992) 2013; S.D. Odintsov, Z. Phys. {\bf C45} (1992) 531;
 I. Antoniadis, P.O. Mazur and E. Mottola, Nucl. Phys. {\bf
B388} (1992) 627.

\bibitem{} I. Antoniadis and S.D. Odintsov, Mod. Phys. Lett.
{\bf A8} (1993) 979; E. Elizalde  and S.D. Odintsov, Int. J.
Mod. Phys. D (1993), to appear.

\bibitem{} E. Elizalde  and S.D. Odintsov, preprint UB-ECM-PF
92/29, to appear in Yad. Fiz. (Sov. J. Nucl. Phys.).

\bibitem{} R. Floreanini, E. Spallucci and R. Percacci, Class.
Quant. Grav. {\bf 8} (1991)  L193; R. Floreanini and R.
Percacci, Phys. Rev. {\bf D46} (1992)  1566.

\bibitem{} R. Floreanini and R. Percacci, preprint SISSA
71/93/EP (1993).

\bibitem{} N. Birrell and P. Davies, {\it Quantum Fields  in
Curved Space-Time}, Cambridge Univ. Press (1982); S. Deser,
M. Duff and C. Isham, Nucl. Phys. {\bf B111} (1976) 45.

\bibitem{} K.S. Stelle, Phys. Rev. {\bf D16} (1977) 953.

\bibitem{} E.T. Tomboulis, Nucl. Phys. {\bf B329} (1990) 410.

\bibitem{} I.L. Buchbinder and S.D. Odintsov, Class. Quant.
Grav. {\bf 2} (1985) 721; E. Elizalde and S.D. Odintsov, Phys.
Lett. {\bf B303} (1993) 240.

\bibitem{} J. Distler and H. Kawai, Nucl. Phys. {\bf B321}
(1989) 509; F. David, Mod. Phys. lett. {\bf A3} (1988) 1651.


\end{thebibliography}
\end{document}